# Computational Framework for Behind-The-Meter DER Techno-Economic Modeling and Optimization—REopt Lite

Sakshi Mishra*†, Josiah Pohl*, Nick Laws*, Dylan Cutler*, Ted Kwasnik*, William Becker*, Alex Zolan*, Kate Anderson*, Dan Olis*, Emma Elgqvist*

*Researcher, Integrated Applications Center, National Renewable Energy Laboratory, Golden, USA
†corresponding author
Emails: Sakshi.Mishra@nrel.gov, josiahpohl@live.com, Nick.Laws@nrel.gov, Dylan.Cutler@nrel.gov, Ted.Kwasnik@nrel.gov, William.Becker@nrel.gov, Alexander.Zolan@nrel.gov, Kate.Anderson@nrel.gov, Dan.Olis@nrel.gov, Emma.Elgqvist@nrel.gov

*Abstract* — The global energy system is undergoing a major transformation. Renewable energy generation is growing and is projected to accelerate further with the global emphasis on decarbonization. Furthermore, distributed generation is projected to play a significant role in the new energy system, and energy models are playing a key role in understanding how distributed generation can be integrated reliably and economically. The deployment of massive amounts of distributed generation requires understanding the interface of technology, economics, and policy in the energy modeling process. In this work, we present an end-to-end computational framework for distributed energy resource (DER) modeling, REopt Lite™, which addresses this need effectively. We describe the problem space, the building blocks of the model, the scaling capabilities of the design, the optimization formulation, and the accessibility of the model. We present a framework for accelerating the techno-economic analysis of behind-the-meter distributed energy resources to enable rapid planning and decision-making, thereby significantly boosting the rate the renewable energy deployment. Lastly, but equally importantly, this computation framework is open-sourced to facilitate transparency, flexibility, and wider collaboration opportunities within the worldwide energy modeling community.

*Keywords* — Renewable Energy Optimization; Mixed-Integer-Linear-Programming; JuMP model; behind-the-meter DERs; microgrid design; open-source energy software; energy modeling

I. INTRODUCTION

### A. Background and Literature Review

U.S. electricity generation from renewable energy sources has doubled since 2008 and currently accounts for about 18% of total generated electricity (Marcy, 2019). This trend is accelerating, with 114 U.S. cities setting their sights as high as 100% renewable energy generation (Herman K, 2019). Worldwide, 28% of electricity was generated from renewable energy sources in 2018 and is projected to reach nearly 50% by 2050 (Bowman, 2019). Renewable sources are expected to surpass all other sources fueling global primary energy consumption by 2025 (EIA, 2019). Furthermore, 30% of the new renewable energy capacity is projected to be decentralized (Henbest, 2017).

Achieving these ambitious deployment targets requires increasingly integrated solutions, where renewable energy is paired with storage and, at times, conventional generation to ensure a sustainable, economically viable, and resilient decentralized energy systems. Moreover, assessing the economic viability of these integrated solutions also requires understanding the temporally dependent complex interactions of energy consumption, resource availability, utility rate tariffs, and policies such as incentives, net metering, and interconnections limits. There are works in the literature that address various aspects of behind-the-meter hybrid distributed energy resource (DER) optimization and modeling. Wu et al. (Wu, Kintner-Meyer, Yang, & Balducci, 2016), model energy storage (battery) as a DER and study the potential benefits and cost savings associated with its deployment and operation. Optimal dispatch alone of battery is studied in (Ngugen & Byrne, 2017) for maximizing savings using peak-shaving and energy arbitrage. Similarly, the value derived from the optimal dispatch of battery systems is modeled and studied in other works, as well. For example, battery dispatch for demand charge reduction is described in (Vatanparvar & Sharma, 2018) and Nguyen et al. tackle power factor correction using a battery system (Nguyen & Byrne, 2018). The sensitivity of the optimal battery operation with respect to various utility tariffs is studied in detail in (Vejdan, Kline, Totri, Grijalva, & Simmons, 2019). Grid-connected photovoltaic (PV) systems alone, on the other hand, are modeled in (Tudu, Mandal, & Chakraborty, 2018).

Integrated optimal operations of battery and PV in the behind-the-meter deployment context has also been studied in the literature. Rockx et al. (Rockx, Tate, & Rogers, 2018) focus on determining the optimal size of the battery and PV system with the aim of reaching grid-parity when net-metering is available. Optimal sizing of a battery in the presence of PV is presented in (Narimani, Asghari, & Sharma, 2018), where optimal sizing is based on the objective of either maximizing the PV utilization or minimizing the demand charge cost. More recently, hybrid energy models incorporating other types of renewable energy resources such as biomass and wind have also been presented (Cuesta, Castillo-Calzadilla, & Borges, 2020). Ringkjob et al. (Ringkjøb, Haugan, & Solbrekke, 2018) present a review of 75 energy modeling tools, that have been developed in various parts of the world for analyzing the integration of renewable energy into various energy systems (both in front of the meter and behind the meter), and Connolly et al. (Connolly, Lund, Mathiesen, & Leahy, 2010) study 37 such tools in greater depth. A review of microgrid-specific optimization of hybrid energy systems is presented in (Fathima & Palanisamy, 2015), whereas (Luna-Rubio, Trejo-Perea, D., & Rios-Moreno, 2012) discusses different methodologies for behind-the-meter hybrid systems in general.

While much of the literature focuses on economic optimization of electric PV and battery systems, less research is available on the behind-the-meter energy modeling problem space where broader electric and heating technologies (PV, wind, combined heat and power, conventional generator, and storage) are modeled in an integrated fashion, coupled with detailed consideration of utility tariff modeling. Additionally, the existing research focuses primarily on the economic optimization of DER and either does not address the potential resilience benefits of these technologies or optimizes the design in a remote location without a connection to the grid (Scioletti, Newman, Goodman, Zolan, & Leyffer, 2017). Our work seeks to fill these two gaps by providing an integrated optimization model that evaluates both the economic and resilience benefits of a broad set of electric and thermal DER technologies.

Furthermore, amongst the works presented in this literature review, few of the models are presented with the capabilities to conduct analysis at scale. In other words, the presented optimization modeling, though very useful for research purposes, is not sufficient to conduct multiple studies efficiently for a wide audience; however, when these optimization models are developed into energy modeling tools, they can have a tangible impact on increasing the real-life deployment of DERs worldwide. This is especially true for behind-the-meter energy modeling tools because they are useful both in the developed and the developing world where the centralized utility grid is not available and localized microgrid solutions are sought after.

Energy modeling tools are utilized at various stages of project implementation to support financial and technical decisions. The insights offered by these tools help system owners, developers, and financiers understand the dynamics between resource availability, energy access, economics, and sustainable development. High-level modeling, or screening, is often performed early in project development to understand whether further investigation is warranted. If initial results are promising, then a more detailed analysis of optimal technology sizes and operation is completed. Examples of free tools that are used to inform project identification include SAM (Blair, et al., System Advisor Model (SAM) General Description, 2018), DER-CAM (Lab), ESyst (Geli), and StorageVET (Institute). HOMER (Lambert, Gilman, & Lillenthal, 2006) is also widely used, though it is not free.

A detailed comparison of these tools and their functionalities can be found in (Krah, 2019). Limitations of some of these tools include[1]: (i) inability to optimally size and dispatch technologies; (ii) inability to consider the integration of multiple technologies; and (iii) lack of transparency. Many of these tools are presented as black-box models; they are available either in the form of a software development kit (SDK) or downloadable desktop application.[2]

Given these limitations, there is still a gap that needs to be bridged to accelerate the large-scale planning and in-depth techno-economic analysis of behind-the-meter DER projects. The work presented here attempts to address that gap by providing a robust, modular, and extensible framework that delivers integrated, optimal energy solutions. By providing a tool that is free, user-friendly, transparent, and scalable we hope to enable large scale analysis and deployment of DER systems.

### B. Relevance, Novelty, and Contribution

The distributed energy implementation decision-making process must consider a range of factors including: (i) technical (physical device constraints, production capabilities, fuel requirements); (ii) economic (capital costs, operation and maintenance costs, utility costs); and (iii) policy (incentives, net metering, interconnection limits). Another layer of complexity is added by site-specific constraints such as land availability, preferred renewable energy technology options, maximum and minimum size specifications, existing DERs, load variability, and critical load requirements. If energy storage is one of the technology options being evaluated, opportunities to optimize the charge and discharge strategy for peak shaving and energy arbitrage further complicate the analysis. The complex nature of the analysis makes it very difficult, if not impossible, to use back-of-the-envelope calculations to find an optimal technology mix that minimizes the life cycle cost of the project while fulfilling the resilience requirements. Traditional metrics such as levelized cost of energy (LCoE) fail to capture the multidimensional interplay of technical, economic, and policy factors and, therefore, do not provide an effective approach for identifying the best solution from the myriad of options available to DER owners. But a comprehensive optimization model capable of considering technological, economic, and policy aspects in the analysis excels in this situation.

---

[1] Note that the described limitations are combined from all the tools, they may or may not be present in every single referenced tool.
[2] SAM being a notable exception, which has been open-sourced.

The framework presented here is designed to answer questions from a broad range of users interested in the optimal sizing, dispatch, and economics of behind-the-meter DER projects. Utilities can use the framework to understand how behind-the-meter systems can provide value to the distribution grid. Commercial site owners can optimize system sizing and controls for maximum value. Project developers can use the model to identify promising market opportunities. Building owners can evaluate which technologies are best for their site. Researchers can evaluate the economic viability of large-scale DERs deployments under varying cost and performance scenarios.

Answering these types of questions requires not just a sophisticated model but also a lot of data. Gathering, curating, and loading the required data (solar and wind resource profiles, utility tariff rates, technology performance, load profiles, and so on) can be a time-consuming task. The framework of REopt Lite enables analysis at scale by automating the collection of most of the data, including resource data, real-life utility tariffs, sample building load profiles, and technology costs, and constructing the input dataset based on them. At the same time, the framework provides the flexibility to adjust default values, to define custom tariffs, and model actual interval data. Furthermore, scenario generation, optimization model creation, solution, and post-processing are also embedded in the automated end-to-end framework. REopt Lite's framework is novel in its ability to make a complex and powerful optimization model with over 100 inputs accessible to professionals without a lot of data or know-how by being able to cost-optimally size and dispatch behind-the-meter DER systems with only four mandatory inputs.

One benefit of this holistic end-to-end optimization system is its relatively easy extensibility. Since the building blocks are arranged in an effectively architected software stack, new state-of-the-art research questions surrounding DER planning can be explored with minor modifications to the inputs, technology modules, and core optimization formulation. Furthermore, the open-source[3] nature of the model allows for transparency, flexibility, and streamlined collaboration. The key contributions of this work are:

- **Optimal, integrated design and dispatch**: An open-source behind-the-meter DER energy modeling framework, based on a mixed integer linear program, utilizing real-life datasets to find optimal sizes and dispatch strategies for multiple integrated technologies;
- **Accessibility**: This framework provides the flexibility to tune over 100 inputs yet requires only four mandatory inputs. This simplifies the analysis for novice users and allows greater use and impact by providing an accessible solution for nonexperts;
- **Automation:** A fully automated end-to-end energy modeling framework enables rapid analyses of thousands of locations and sensitivities
- **Extensibility**: An extensible framework provides the capability to incorporate additional decision variables and constraints, providing a foundation to address future research questions. These could include analysis of other technologies in the smart buildings and district systems modeling space, such as flexible load modeling and electrical vehicle charging station models.

REopt Lite's main contributions include its sophisticated and modular software stack that enables automated data fetching, preprocessing, optimizing, and post-processing steps, making it an effective out-of-the-box energy modeling computational tool for subject-matters experts and nonexperts alike.

*C. Article Organization*

With the background, motivation, and contributions laid out in the previous subsections, the rest of the article is organized as follows. Section II describes the building blocks of REopt Lite's modularized input data and data sources. Section III delineates the core modeling framework of REopt Lite. Section V introduces potential future extensions to answer state-of-the-art research questions within this framework. The article is concluded in Section VI with a discussion of future research directions and planned framework enhancements.

---

[3] Open-sourced under BSD-3 License, which is permissive with minimal restriction on use and distribution.

II. PROBLEM FORMULATION AND BUILDING BLOCKS

In this section, we first describe the overall problem formulation of REopt Lite. We then describe six core building blocks of the model: PVWatts®, Wind Toolkit, SAM, URDB, DOE Commercial Reference Buildings, and AVERT. REopt Lite calls the Application Programming Interfaces (APIs) of these tools in an automated fashion to obtain accurate site resource, generation, utility rate tariffs, load datasets, and grid $CO_2$ emissions; therefore, we refer to them as the building blocks of REopt Lite.

### A. Problem Formulation

Formulated as a mixed-integer linear program, REopt Lite solves a deterministic optimization problem to determine the optimal selection, sizing, and dispatch strategy of technologies chosen from a candidate pool such that electrical and thermal loads are met at every time step at the minimum life cycle cost. REopt Lite is a time series model where energy balances are ensured at each timestep (typically 1-hour intervals) and operational constraints are upheld while minimizing the cost of energy services for a given customer. A primary modeling assumption is that decisions made by the model will not impact the markets (i.e., the model is always assumed to be a price-taker). This is in contrast to unit commitment and dispatch models where pricing is a decision variable. REopt Lite does not model power flow or transient effects.

REopt Lite solves a single year optimization to determine N-year cash flows, assuming constant production and consumption over all N years of the desired analysis period. The model minimizes total life cycle cost, comprising a set of possible revenues and expenses, over the analysis period subject to a variety of constraints to ensure thermal and electrical loads are met at every time step by some combination of candidate technologies. Costs considered include capital costs; operating expenses like utility purchases, fuel costs, and operation and maintenance costs; operating revenues like production-based incentives or net metering income; and capital cost incentives. Cash flows during the analysis period are found by first escalating the present costs at project-specific inflation and utility cost escalation rates, then discounting back to the present. It assumes perfect prediction of all future events, including weather and load.

REopt Lite offers two types of analysis: (i) Financial and (ii) Resilience. In the financial analysis option, REopt Lite finds the optimal system design and dispatch that minimizes the life cycle cost of energy. The resilience analysis option does the same thing, but with the added constraint that the system must be able to sustain the critical load without the utility grid during a user-specified outage. While a financial analysis will always provide a net present value greater than or equal to zero, the net present value of a resilience analysis may be negative if the assets required to meet the critical load for the outage duration are not economically viable. The main inputs and outputs of the model are described in Figure 1. REopt Lite considers three classes of drivers (goals, economics/ policy, and utility costs), to provide a set of optimal outputs (system design, operations, and economics). These outputs help decision makers assess the technical and financial viability of a project.

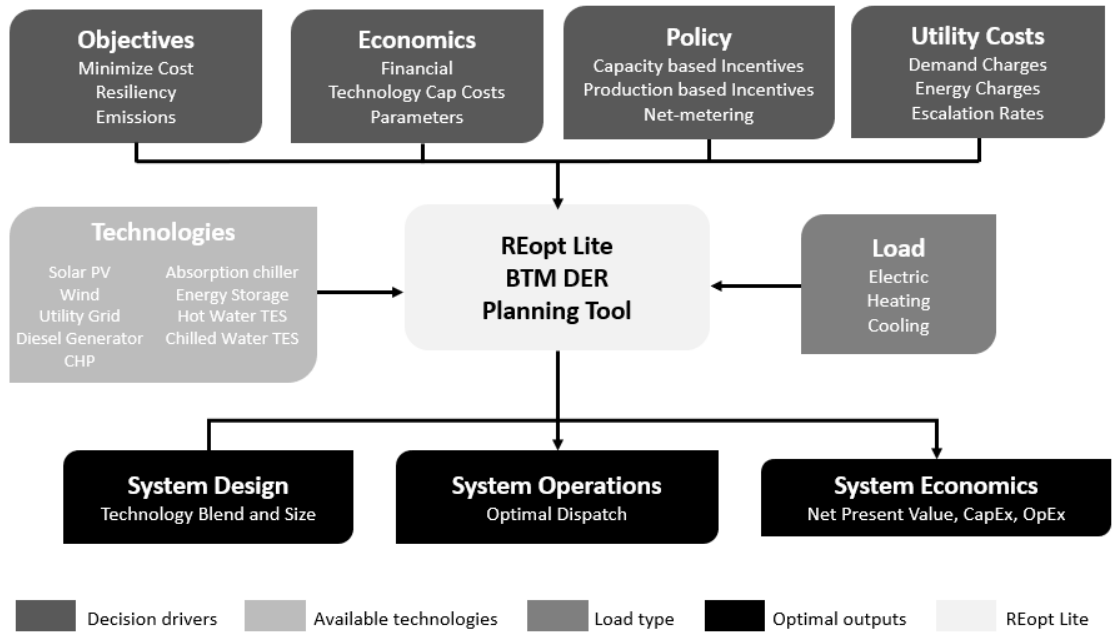

Figure 1. REopt Lite: Inputs and outputs[4]

### B. PVWatts—Solar Resource and Production Data

PVWatts is an NREL-developed tool that estimates the energy production and cost of grid-connected PV energy systems (NREL, PVWatts Calculator, 2020) (Dobos, 2014). PVWatts uses typical meteorological year (TMY) data from the NREL National Solar Radiation Database (NSRDB) to estimate PV energy production based on multiyear averages for U.S. and international locations. REopt Lite uses the location information provided by the user to query the PVWatts API for solar production factors based on the available solar resource at the given location.

### C. Wind Toolkit—Wind Resource Data

The Wind Integration National Dataset (WIND) Toolkit provides wind resource data for over 126,000 sites in the continental United States (Draxl, Clifon, Hodge, & McCaa, 2015). REopt Lite utilizes the Wind Toolkit to obtain wind speed, air pressure, air temperature, and wind direction at an hourly resolution based on the geographic location of the site.

### D. System Advisory Model—Wind Production Data

SAM is a simulation-based techno-economic model that evaluates the technical and financial performance of renewable energy projects. (Blair, et al., System Advisor Model (SAM) General Description, 2017). In REopt Lite, SAM is utilized to produce wind generation data based on the Wind Toolkit wind resource data and wind turbine power curves for representative turbine sizes (large, medium, commercial, and residential). The wind generation data is translated into a wind production factor (energy generated per unit of system capacity) for use in REopt Lite.

### E. URDB—Utility Rate Tariff Data

The Utility Rate Database (URDB) is a publicly accessible repository of over 52,000 U.S. and international utility rates (OpenEI, 2017). The URDB captures time-of-use and tiered structures for both energy and demand charges, as well as fixed charges. REopt Lite can select a rate from the URDB by its label, a unique identifier. REopt Lite can also process a user-defined custom rate, so long as the custom rate has the same structure as a typical URDB response.[5]

---

[4] Full forms for the acronyms used in this figure: (i) CHP–combined heat and power; (ii) TES–thermal energy storage; (iii) Cap—capital; (iv) CapEx–Capital expenditure; (v) OpEx–operational expenditure.
[5] See https://openei.org/services/doc/rest/util_rates/?version=7#response-fields.

### F. Load Profiles—DOE Commercial Reference Buildings

REopt Lite provides default load profiles based on common commercial building types obtained from the DOE Commercial Reference Buildings data set (Energy, n.d.). There are 240 unique reference load profiles built from detailed building energy models of 16 building types in 16 US climate zones.[6] These represent 70% of the national commercial building stock (Deru, et al., 2011). This data set is stored locally within the API. There are three load profiles used from the Commercial Reference Buildings data set including total electric load, heating boiler fuel load, and the electric load attributed to cooling (chilled water). The electric load attributed to cooling is a subset of the total electric load. The user can also upload their own interval data.

### G. Emissions—AVERT Database

The $CO_2$ emissions analysis follows the general methodology and guidance described in the U.S. Environmental Protection Agency's (EPA's) Fuel and Carbon Dioxide Emissions Savings Calculation Methodology for Combined Heat and Power Systems (EPA, Fuel and Carbon Dioxide Emissions Savings Calculation Methodology for Combined Heat and Power Systems, 2015). For sites inside the continental United States, we employ regional hourly emission factors from EPA's AVERT tool (EPA, AVoided Emissions and geneRation Tool (AVERT), 2019). For sites in Alaska and Hawaii, we apply EPA's eGRID annual average emission factors, as described in (EPA, Fuel and Carbon Dioxide Emissions Savings Calculation Methodology for Combined Heat and Power Systems, 2015). For locations more than 5 miles outside the boundary of Alaska, Hawaii, or an AVERT region, we do not assign an emission factor but allow the user to enter their own.

## III. MODELING FRAMEWORK

In this section, we describe REopt Lite's core modeling framework. First, we review the overall software architecture including the outermost API and task-management modules that enable programmatic access and analysis at scale. Then we describe the core modules, including the python layer for technology modeling and data preprocessing (where techno-economic models are created) and optimization layer.

### A. End-to-End Computational Pipeline

REopt Lite's modeling framework has six major components: (1) API; (2) Task management; (3) Techno-economic models; (4) Optimization; (5) Post-processing; and (6) Resilience outage simulator. Although the API and Task management modules in REopt Lite serve important purposes, they have mainly been adapted from available python libraries and weaved together to function in tandem. The novel contributions of REopt Lite, from a research contribution perspective, come from its core modules (outlined with dashed boundaries in Figure 2). These modules are presented in detail in the following subsections.

---

[6] See https://www.energy.gov/eere/buildings/commercial-reference-buildings.

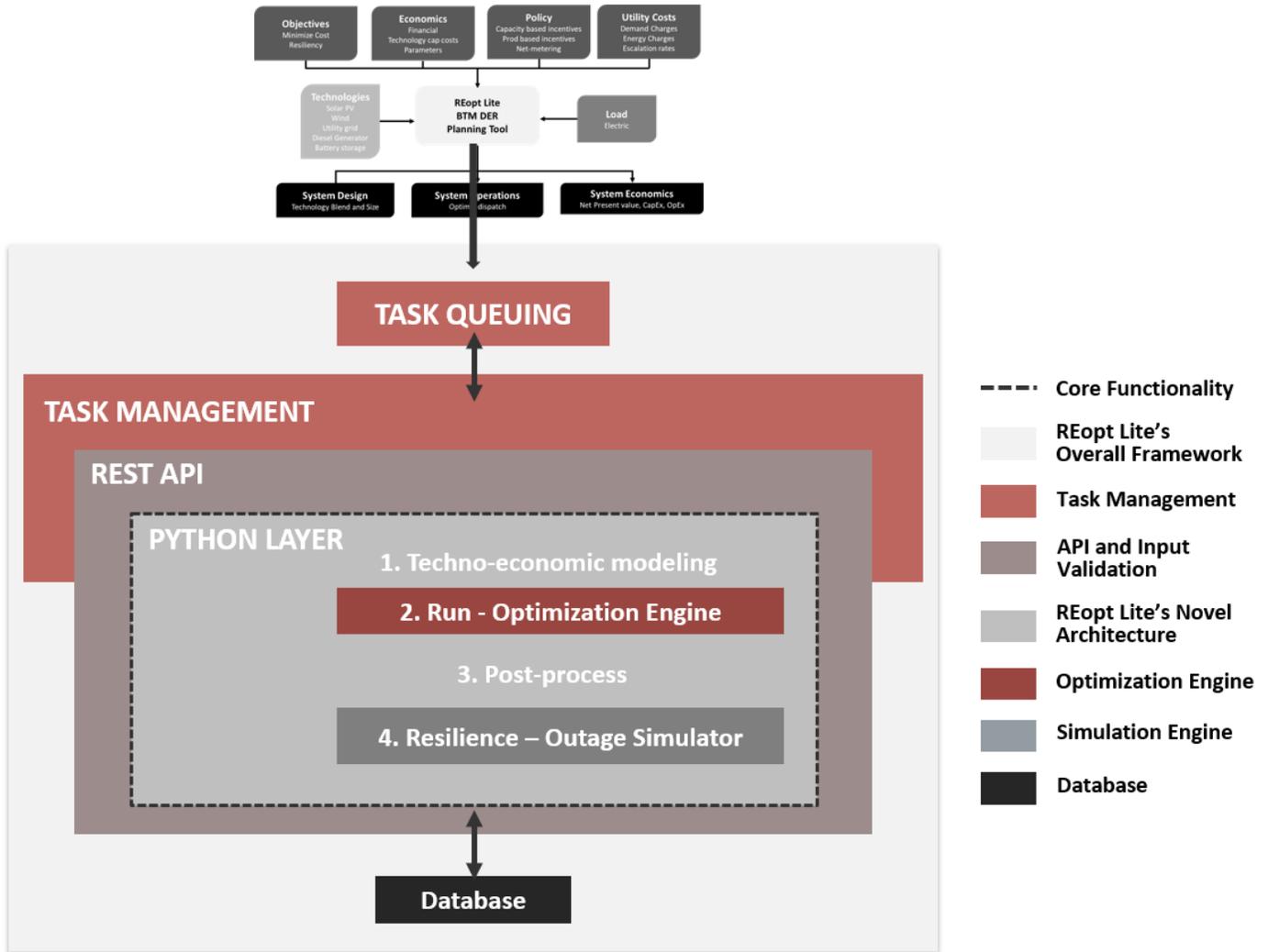

Figure 2. REopt Lite: End-to-end computation pipeline

B. *API and Task Management* [7]

REopt Lite's API module allows developers to write a program that requests a scenario run from its servers programmatically (the server can be a local host or NREL hosted REopt Lite servers[8]). Other functionalities of the API module include validation of the input posts (along with exception handling) and saving inputs/outputs to the Postgres database. When an analysis request is sent in the form of a JSON input file, it is first processed through the API module. After this, the validation task management module facilitates the sequential execution of the technology creation module, followed by the optimization module, and then the post-processing module. The optimization module is run twice—once for the business-as-usual case and a second time for the optimal design case (referred to as optimal case). The task management module facilitates the parallel execution of these two optimization-runs for a single scenario, where task queueing is a sub-task. Scenario.py, (business as usual) src/reopt.jl, (Optimal) src/reopt.jl, and process_results.py are the four separate tasks that celery manages to execute in parallel, as shown in Figure 3. Similarly, it also enables the handling of thousands of scenarios, when posted at the same time, by distributing the jobs to the available compute resources while maintaining the pre-assigned sequence.

---

[7] Though REopt uses Django, Postgres, Celery, Redis for API, data storage, and task management (ordering and queuing), respectively, there are other open-source packages available to accomplish the same objectives. NREL does not endorse any specific package.
[8] https://developer.nrel.gov/docs/energy-optimization/reopt-v1/.

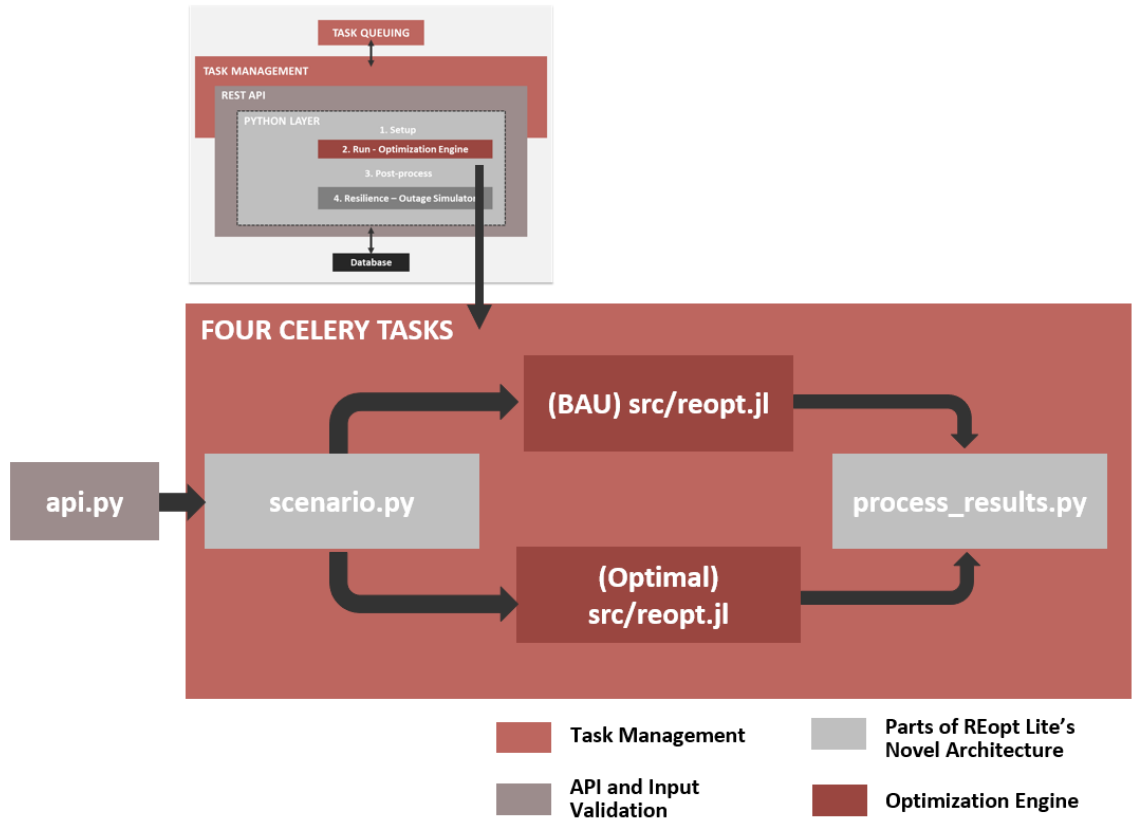

**Figure 3. Task management**

*C. Techno-Economic Models*

The techno-economic modeling is performed in the technology creation module that precedes the core optimization layer (discussed in Section III.D). This module contains our framework's novel architecture (dashed boundaries in Figure 2), where the majority of the modeling takes place. Generation technologies, storage, and load are modeled and converted into the multidimensional matrices that are then provided as inputs to the optimization module. Also, the utility tariff (user-defined or fetched from URDB) is parsed and converted into single-dimensional arrays. Figure 4 shows the algorithmic flow of the module where the following steps are executed sequentially: (i) type of analysis selection; (ii) mandatory site-specific inputs; (iii) automated data fetch and curation process; (iv) techno-economic modeling in the technology creation module. The type of analysis selection and mandatory inputs are explained in detail in (NREL, REopt Lite - Webtool User Manual).

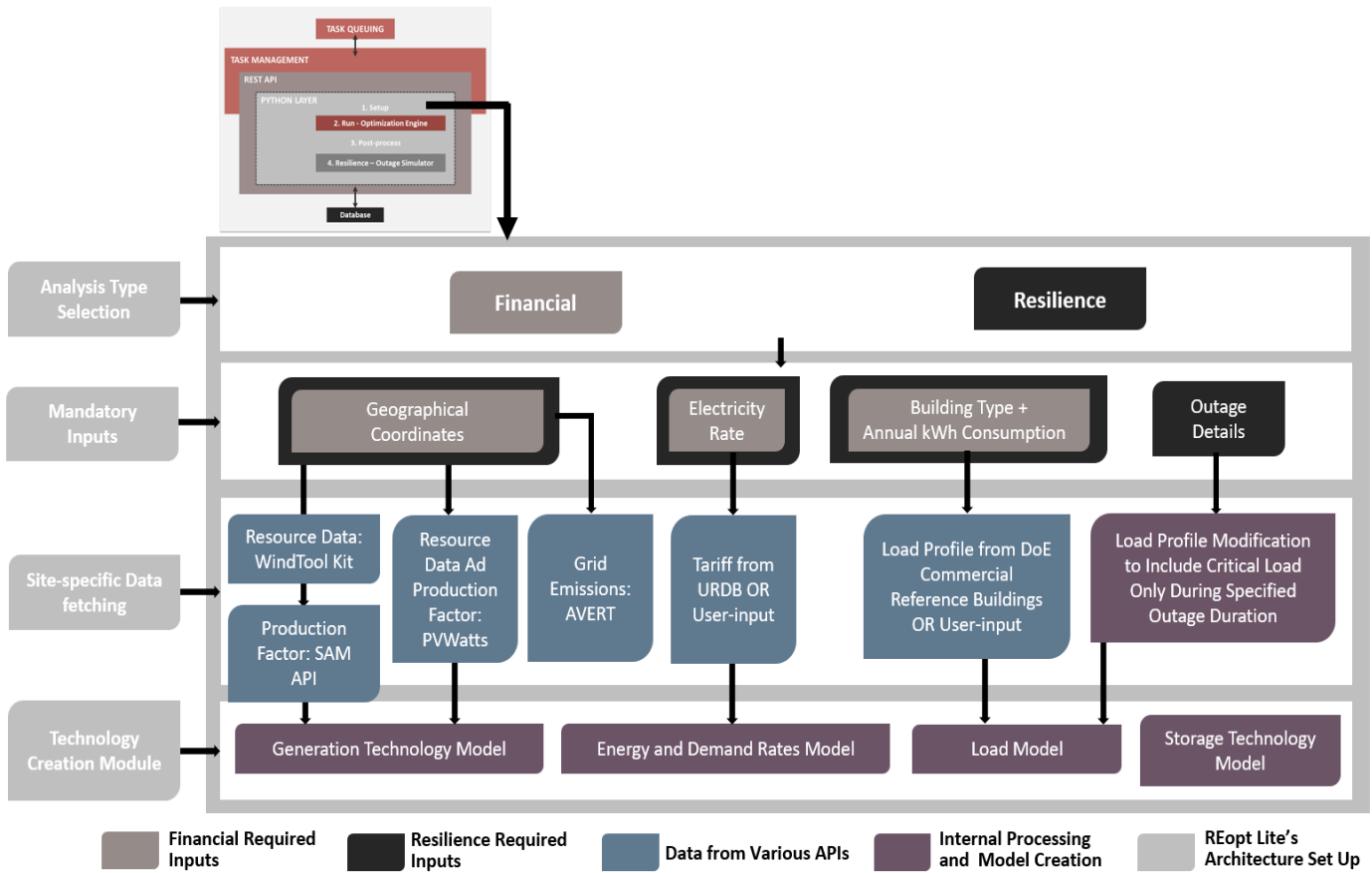

**Figure 4. Techno-economic modeling**

*1) Generation Modeling*

There are four DER electric generation technologies modeled in the framework: (i) PV; (ii) Wind; (iii) Combined heat and power (CHP); and (iv) Diesel Generator.[9] The fifth generation source is the utility itself. For PV and wind, a location-specific time-series production factor is calculated based on the available resource using PVWatts (PV) or the Wind Toolkit and SAM (wind). The production factor represents the amount of energy that can be generated per unit of installed capacity in each timestep. CHP produces both electric and thermal energy for the site. For CHP, site location attributes such as elevation and outdoor air temperature affect the electric capacity and efficiency of the prime mover. Part-load electric efficiency and heat recovery performance are described in more detail in (Becker, Cutler, Anderson, & Olis, 2019). An absorption chiller that produces chilled water from a supply of hot thermal energy may also be considered in conjunction with CHP. The production factor of the diesel generator is zero for all hours of the year except during the user-specified outage period—essentially modeling it as a backup generation resource. Utility supply is modeled as an infinite source of energy for the site, which is turned off only for the outage duration.

The concept of "Tech" and "TechClass" in REopt Lite's generation modeling is somewhat unique. It facilitates an effective way to model a single type of generation technology with two variations—with and without net metering—allowing the optimization module to consider net metering as part of the mathematical formulation. "Tech" is an array of generic technology that can meet the load(s). All techs can have capital costs, fuel consumption (entered as zero for PV and Wind), and operation and maintenacne costs (fixed as well as variable). Techs can also have production- and capacity-based incentives and tax benefits. "TechClass" is an abstract group of "Techs" used for modeling the constraints associated with net-metering. For example, there are two sizing choices for a PV system: (i) PV eligible for net metering benefits, that cannot generate more than the annual site consumption; and (ii) PV not eligible for net metering that can generate an "unlimited" amount of energy (only constrained if a maximum system size is

---
[9] Diesel generator is restricted to resilience applications only in the webtool; grid-connected use of diesel generators is allowed in the API.

provided[10]). The same concept applies to the wind system model. A matrix named "TechToTechClassMatrix" associates "Techs" with "TechClass", which is explained further in mathematical formulation presented in (Ogunmodede, Anderson, Cutler, & Newman).

### 2) *Storage Modeling*

Battery storage is modeled as a "reservoir" in REopt, wherein electric energy produced during one time step can be consumed during another. The model imposes heuristic constraints to account for characteristics of the battery, such as round-trip efficiency, minimum state of charge, and charge/discharge rate limits. The model independently determines the optimal size of the energy capacity of the battery in kWh and the power delivery (inverter) in kW. By default, any electric-generating technology can charge the battery, but charging can also be limited to specific technologies.

Two types of thermal energy storage (TES) are also considered in the form of hot water and chilled water storage tanks. The tanks store thermal energy to decouple production from consumption. The optimization models for hot and chilled water are the same; only the default cost and performance parameter values are different. TES is modeled similar to battery storage but with the addition of a thermal decay term, which captures the time-dependent aspect of useful heat loss from the tank. The hot water TES interfaces with the boiler, CHP, and the heating load. The chilled water TES interfaces with the electric chiller, the absorption chiller (if considered), and the cooling load.

### 3) *Load Modeling*

The electric load profile is generated by several possible sets of inputs:
1) The reference building type and location (for climate zone)
2) The reference building type, location, and user-defined aggregated annual or monthly load(s)
3) User-defined hourly or 15-minute load profile.

Option 1 uses the default DOE Commercial Reference Building load profiles based on the building type and location. Option 2 applies the normalized reference building load profile to create a load profile that matches user-entered aggregated annual or monthly loads. Both Options 1 and 2 leverage the *geopandas* python package to determine the nearest city on which 16 climate zones are based. In Option 3, the user inputs hourly or 15-minute interval data for the electricity load; this option does not utilize the DOE Commercial Reference Buildings or location-based climate zones. The framework also allows for building hybrid load profiles where more than one DOE-reference building type can be specified, and a weighted average of the typical building-type based load profiles is calculated to build a campus-wide load profile. The user can similarly define the critical load for a resilience analysis based on either a percentage of the typical load or a user-defined load profile.

The Commercial Reference Buildings database is also leveraged for heating and cooling loads. The flow of energy from generation to storage and load is shown in Figure 5. The heating load has identical input options as the electric load profile. The cooling load has different options for creating a custom cooling load profile relative to the total electric load. When electric chillers are included in technology option being assessed, the model determines the dispatch of the electric load's portion that is served by chillers, encircled (elec. chiller load) in Figure 5. The user can specify a fixed-annual or monthly-varying fraction of the total electric load to be used for the cooling electric load profile. The heating load is the site's original heating boiler fuel load, and this can be met by the existing boiler, CHP, or hot water TES. If the absorption chiller is considered, its hot thermal load is added to the original heating load in the model optimization, and this new heating thermal load must be met by the same technologies (encircled in Figure 5 as absorp. chiller load). Since the cooling load (electric load from the electric chiller) is a subset of the total electric load (retail load), there is not a separate load category for the cooling load. The cooling load electric portion of the total electric load must be met by the electric chiller, the absorption chiller, or discharge from the chilled water TES.

---

[10] Theoretically unlimited, but in the code implementation a big number (10e-7) is used.

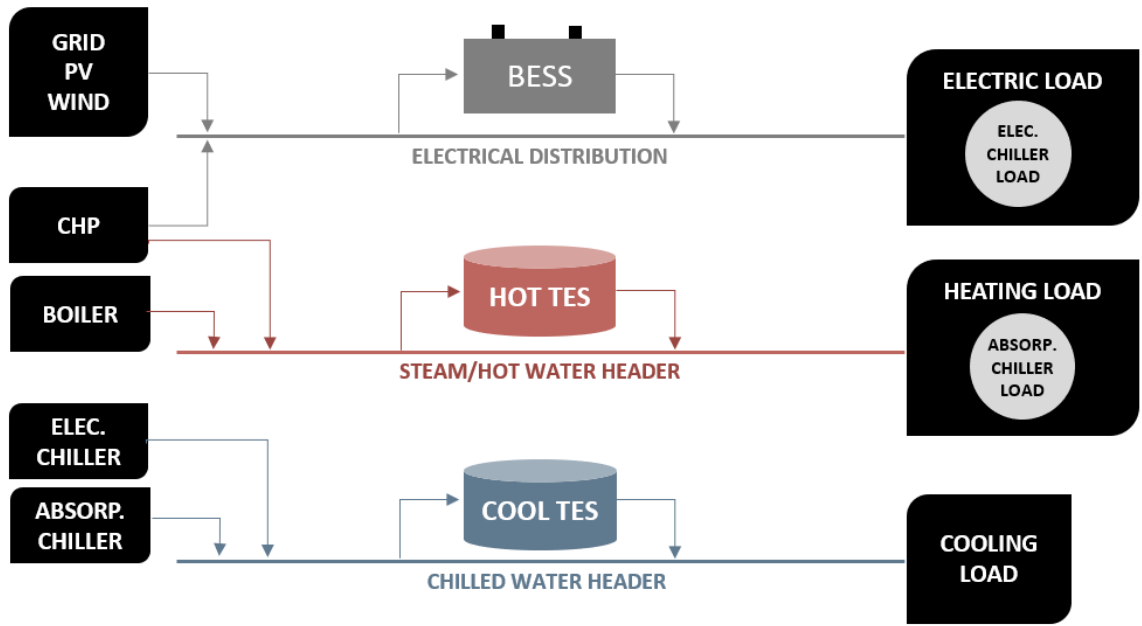

Figure 5. REopt Lite's generation, storage, and load categories

*4) Energy and Demand Rates*

REopt Lite models both simple and complex utility tariff inputs. At a minimum, blended annual or monthly energy and demand rates can be submitted to the model. Alternatively, the model accepts detailed utility tariffs that comply with the URDB v7 format.[11] This more complex format captures energy and demand charges with seasonal variability across different months, time-of-use components (variability assessed over hourly periods) and tiers (charges that are differentiated based on the magnitude of energy or power use). It also includes fixed monthly charges, minimum annual or monthly charges, and demand charge lookbacks. REopt Lite accepts complete URDB v7 utility tariff descriptions and can also look them up using a unique label if they exist in the URDB v7. Hourly real-time prices can also be uploaded.

The net metering input determines the maximum size of total systems that can be installed under a net metering agreement with the utility. Projects sized up to the net metering limit are modeled to receive credit for any exported energy at the electric retail rate at the time of export.[12] Projects sized greater than the net metering limit will receive credit at the wholesale rate for any energy exported. Projects not under a net metering agreement receive the wholesale rate for all exported energy, up to the annual site. Energy exported in excess of the annual site load receives zero credit.

*D. Optimization Module—Design Principles*

The techno-economic modeling module generates preprocessed data, in the form of matrices, that are fed to the optimization model at the core of the computational framework. The following subsections present a conceptual description of the formulation, including the objective and constraints. The full mathematical formulation, including equations for the optimization models, is presented as part of the documentation in the GitHub repository,[13] where the code is made publicly available.

*1) Model formulation—Mixed Integer Linear Programming*

REopt Lite's mathematical optimization model is formulated as a Mixed Integer Linear Program in a modeling language called JuMP (Dunning, Huchette, & Lubin, 2017), which is embedded in the Julia programming language. The objective of the techno-economic optimization problem is to minimize the life cycle cost of energy, from the perspective of a single decision-maker (the behind-the-meter DER owner). It uses 1 year of resource and cost data with present worth factors to account for lifetime costs. It assumes that 1 year repeats with degradation and escalation factors. Energy balance is maintained at every time step for the entire year by ensuring the load is met from some combination of grid purchases, on-site generation, or discharge from energy storage.

---

[11] https://openei.org/services/doc/rest/util_rates/?version=7.
[12] Information on state net metering limits is available at https://www.dsireusa.org/.
[13] https://github.com/NREL/REopt_Lite_API/wiki/REopt-Mathematical-Model-Documentation.

Currently, the model does not include power flow or transient effects and has a perfect prediction about upcoming weather and load events.

The model finds the optimal technology sizes (which can be zero) and optimal dispatch strategy to minimize the life cycle cost of energy subject to resource, operating, and market constraints. Binary and continuous decision variables are used for sizing decisions in the model—binary to decide whether a technology should be installed, and continuous to decide the optimal size of the technology being recommended for installation.

*2) Nature and Types of Constraints*

We categorize the types of constraints in REopt Lite into: (i) design; (ii) operational; and (iii) policy, as shown in Figure 6.
- The design constraints include real-life site-specific constraints such as land area availability and maximum fuel availability (depending on the fuel storage capacity of the site). They also enable users to specify the desired minimum and maximum technology size.
- The operational constraints maintain the generation and demand balance at all time steps and allocate the load into the correct demand and energy tiers of the utility tariff. They also represent the physical limits of the generators (i.e., minimum turndown).
- Last, policy constraints model the rules of incentive programs like investment tax credits, rebates, renewable energy credits, and net metering. The maximum value of production and capacity-based incentives and the net metering limit are incorporated into a dedicated set of constraints.

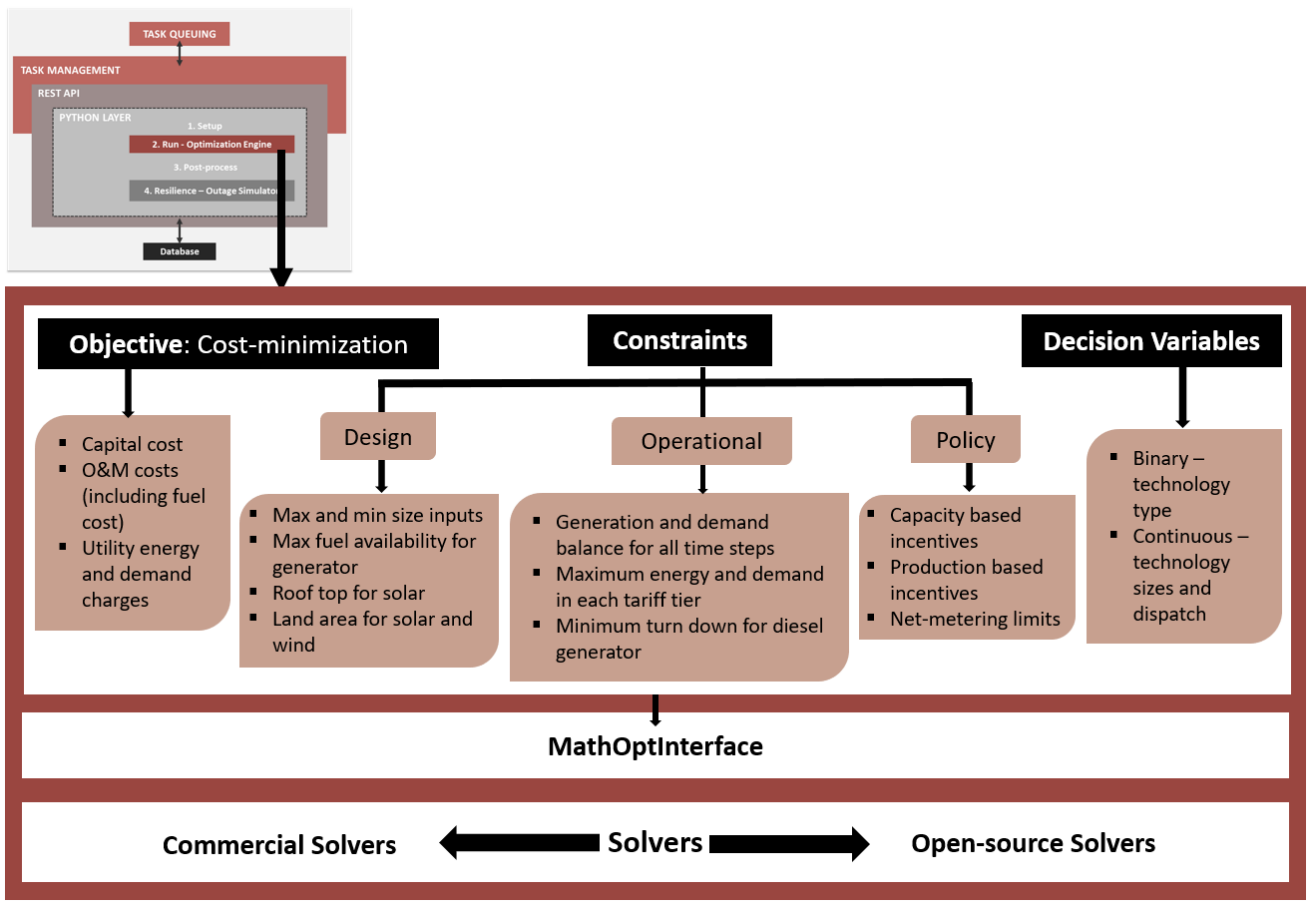

Figure 6. Overall optimization formulation with various layers

*3) Solvers*

Using the metaprogramming feature of the Julia language, JuMP provides a set of macros that specify variable, constraint, and expression definitions. These macros are unpacked into code that declares MathOptInterface typed variables that will be passed onto the solver. MathOptInterface creates a solver agnostic representation of the monolithic optimization problem in standard form.

This standard form is converted into the solver specific types and arranged in the form of julia's composite datatypes (called structs) using the corresponding Julia package (i.e. *Cbc.jl, Xpress.jl*). These packages convert the MathOptInterface representation into the format required by the solver-specific API. Once the data is input through the respective API, the solver's shared libraries are used to perform the optimization computations. Using JuMP allows REopt Lite to interface with any solver that has a Julia package which includes commercial solvers Xpress, CPLEX, Gurobi, and open-source solvers CBC, GLPK, and SCIP along with many others.

### E. Result Post-Processing Module

The final step in the REopt Lite workflow is a python script for processing the results. This step takes the outputs from the two parallel optimization models (financial and resilience) and combines them into a nested dictionary with high-level keys corresponding to each of the objects in the modeling framework. The nested dictionary mirrors the JSON structure required for POST'ing jobs to the API, as well as the API response structure. Each of the high-level keys, such as "PV," "Wind," and "Financial" correspond to a table in the database where the inputs and outputs of the API are saved.

The post-processing module also runs the outage simulator to calculate the statistics around outage survival rates (described in the next section). Parameters that depend on both the BAU and optimal scenarios are also calculated in the post-processing module, such as the net present value and energy bill savings.

### F. Resilience Outage Simulation Module

As shown in Figure 2, the outage simulator is a part of the python layer, which is executed after the optimization module. The function of this module is to simulate outages starting every hour of the year, and then evaluate the number of hours the system is able to sustain the critical load while the utility supply is unavailable. This is different than the optimization model, which optimized the system to sustain the load for one specific outage period. The outage simulator evaluates how that system would perform during outages occurring at other times of the year. The outage simulator only considers the one system design (i.e., types and sizes of technologies) recommended by the optimization module.

The outage simulator obtains the battery state of charge at each hour of the year from the optimization module, and the starting battery state of charge for each outage simulation corresponds to the state of charge in the corresponding hour of the optimal dispatch strategy. The simulator calculates the probabilities of survival annually, as well as by outage start month and outage start hour. In this way, the outage simulator adds another layer of resilience modeling to this computational framework by offering probabilistic insights into the resilience performance of the system over the full year.

### G. Endpoints of the Framework

Various API endpoints enable information retrieval from the tool. Having multiple endpoints allows for greater control over structured information exchange with the API. Except for the first end point available for REopt Lite API,[14] all other end points are GET requests that retrieve results and associated information from the API. New optimization tasks/analysis with parameters formatted as a JSON are submitted to the first endpoint as POST requests.[15] If the posted JSON passes validation, the JSON response from the endpoint simply contains a unique identifier for the task (a *run uuid*). This run uuid is then used for creating the URLs that submit the GET request to the rest of the endpoints.

---

[14] i) v1/job/
 ii) v1/job/<run_uuid>/results
 iii) v1/job/<run_uuid>/proforma
 iv) v1/help
 v) v1/annual_kwh?doe_reference_name=<name>&latitude=<lat>&longitude=<long>
 vi) v1/simulated_load?doe_reference_name=<name>&latitude=<lat>&longitude=<long>&annual_kwh=<value>&monthly_totals_kwh=<list of value>
 vii) v1/ generator_efficiency?generator_kw=<value>
 viii) v1/user/<user uuid>/summary
[15] See https://developer.nrel.gov/docs/energy-optimization/reopt-v1/ for valid POST format.

## IV. EXTENSIBILITY OF THE FRAMEWORK

REopt Lite covers a wide collection of economic incentives, utility pricing structures, operating modes, and technology combinations for a prospective DER design; however, there exist myriad potential extensions of the tool, which include but are not limited to:
- providing ancillary services
- optimizing DER design and dispatch decisions under load and renewable resource uncertainty
- adding power flows to multiple sites
- including critical and flexible loads with varying penalties for a shortfall in meeting loads
- introducing additional site loads to conduct unified analyses of the energy, thermal, and water system.

In what follows, we illustrate the extensibility of the framework using the first example above as a guide for extending the model. We describe the specific ancillary service that we intend to provide and then propose an extension of the existing framework to cover the service.

### A. Ancillary Services Overview

Beyond the dispatch and sales of energy to the grid at wholesale or net metering rates, some DERs have the ability to provide ancillary services to the grid. Hirst and Kirby (Hirst & Brendan, 1996) describe ancillary services in detail, which include but are not limited to:
- secondary control, or the capacity to respond to either rapid changes in load or the sudden outage of a generator
- voltage and frequency regulation, or the correction of electrical imbalances that might affect the stability of the power system
- black-start regulation, or the ability to supply power during system restoration after a grid outage.

Rebours et al. (Rebours, Kirschen, Trotignon, & Rossignol, 2007) describe existing markets for producers to provide these services to the grid, and Hubert and Andersson (Abgottspon & Andersson, 2012) propose a model that obtains optimal energy dispatch and ancillary service bids for a hydroelectric generator as a two-stage stochastic program. The DER designs in REopt Lite have the option to "island" from the grid to meet site load during a grid outage, which may motivate an extension in which secondary control within the microgrid must be maintained at all times; therefore, in our example, we propose an extension in which spinning reserves for secondary control must be maintained at some prespecified level in each time period.

### B. Model Extension: On-Site Spinning Reserve Requirement

The notation that follows in this sample uses lowercase or Greek letters for parameters and capital letters for decision variables; superscripts denote descriptors, while subscripts denote indices on sets. In this extension, we assume that like the load, the spinning reserve requirement in each period is deterministic. Additionally, we assume that the required spinning reserve is bidirectional, and may only be covered by storage and running dispatchable technologies (i.e., PV and wind generators are excluded). Let $h \in H$ be the collection of time periods in the proposed DER's operating horizon and let $\delta_h^{sr}$ be the spinning reserve requirement in period $h$. The downward spinning reserve requirement, or the flexibility to decrease produced power can be modeled via the following constraint:

$$\sum_{t \in T^{td}} f_{th}^p X_{th}^{rp} + \sum_{b \in B^e} \min\{X_b^{bkW}, (X_b^{bkWh} - X_{bh}^{se})/\Delta\} \geq \delta_h^{sr}, \forall h \in H,$$

in which REopt Lite has already defined the following parameters, sets, and variables as follows:
- $T^{td}$ is the collection of dispatchable, electricity-producing technologies;
- $f_{th}^p$ is the production (i.e., derate) factor of technology $t$ in period $h$;
- $X_{th}^{rp}$ is the rated production of technology $t$ in period $h$ (in kW);
- $B^e$ is the collection of electrical storage technologies;
- $X_b^{bkW}$ is the battery's power rating (in kW);
- $X_b^{bkWh}$ is the battery's energy rating (in kWh);
- $X_{bh}^{se}$ is the battery's state of charge in period h (in kWh); and,
- $\Delta$ is the period length (in h).

Likewise, the upward spinning reserve requirement can be modeled as the following constraint:

$$\sum_{t \in T^{td}} (Z_{th}^{to} \cdot X_t^{\sigma} - f_t^p X_{th}^{rp}) + \sum_{b \in B^e} \min\{X_b^{bkW}, X_{bh}^{se}/\Delta\} \geq \delta_h^{sr}, \forall h \in H,$$

in which REopt Lite has defined the following additional variables:

- $X_t^\sigma$ is the power rating of technology $t$ (in kW); and
- $Z_{th}^{to}$ is 1 if technology $t$ is turned on in period $h$, and 0 otherwise.

REopt Lite's mixed integer linear programming structure precludes the inclusion of the product of a binary and continuous variable in the formulation, so we can implement the substitution:

$$Y_{th} = Z_{th}^{to} \cdot X_t^\sigma, \forall t \in T^{td}, h \in H,$$

and reformulate the upward spinning reserve constraint via the following:

$$\sum_{t \in T^{td}} (Y_{th} - f_t^p X_{th}^{rp}) + \sum_{b \in B^e} \min\{X_b^{bkW}, X_{bh}^{se}/\Delta\} \geq \delta_h^{sr}, \forall h \in H$$

$$Y_{th} \leq X_t^\sigma, \forall t \in T^{td}, h \in H,$$
$$Y_{th} \leq b^\sigma Z_{th}^{to}, \forall t \in T^{td}, h \in H,$$
$$Y_{th} \geq X_t^\sigma - b^\sigma \cdot (1 - Z_{th}^{to}), \forall t \in T^{td}, h \in H.$$

Aside from the additional constraints, implementing this within the REopt Lite framework only requires the addition of a single time-series parameter, $\delta_h^{sr}$, and a new auxiliary variable, $Y_{th}$, which demonstrates the flexibility of the existing framework to handle new extensions with limited development time.

## V. FUTURE WORK

This framework can also be extended to answer other research questions in the energy modeling domain. The modular pipeline allows adding functionalities with a relatively low programming burden. Some examples include:

- The integrated energy modeling space offers tremendous possibilities for enhancing efficiencies by leveraging the integration of energy, thermal, and water systems to conduct unified analyses. REopt Lite's computation framework, designed to be easily extensible, has the potential to grow as an integrated energy modeling framework.
- The framework can be further developed to incorporate real-time user preferences and controls for operating the intelligent grid-edge devices in the optimal dispatch of the technologies, loads, and storage.
- The framework can also be used to study the role of behind-the-meter resources in supporting the reliability and resilience of the distribution grid. The participation of these assets in energy markets can also be studied by combining REopt Lite's capabilities with distribution system modeling tools (Petrovic, Strezoski, & Dumnic, 2019).
- Currently, REopt Lite models resilience in a deterministic paradigm where the assets are sized to sustain the critical load for the defined outage period. The resilience simulation module then estimates the probabilities based on the fixed optimal asset size recommended by the optimization module. The resilience modeling of the proposed framework can be further built upon to consider uncertainty associated with the outages.
- Flexible loads can be modeled within REopt Lite's computation framework to support the Grid-Interactive Efficient Buildings initiative of DOE (Neukomm, Nubbe, & Fares, 2019). When paired with the thermal modeling, flexible load modeling can offer considerable efficiency gains and cost minimization for the buildings.
- The REopt Lite framework can be extended to analyze the impacts of electric vehicle interactions with the grid. This could include conducting cost-benefit analysis of DC fast-charging applications and determining the value of using parked electric vehicles for demand management applications.

## VI. SUMMARY AND OUTLOOK

The proposed computational framework, REopt Lite, is a robust and extensible energy modeling tool for conducting behind-the-meter DER analysis. We present the detailed algorithmic flow and code implementation of the framework, along with its interactions with the building blocks that serve as the data sources of the optimization formulation. We also present a modular and comprehensive component modeling approach as well as the overall mixed integer linear programming-based optimization formulation concept employed by the computational framework. Extensive emphasis is placed on automation, accessibility, and extensibility in the end-to-end framework design with the goal of making an effective behind-the-meter energy planning tool useful for subject matter experts and novice users alike. Moreover, owing to the versatility of the framework, there are many opportunities for extending this framework to answer a variety of research questions surrounding the smart grid domain.

The presented framework, though useful, has limitations that need to be addressed to enhance its effectiveness and capabilities. Currently, one of the major limitations of this framework is its deterministic approach. Extension to a stochastic optimization formulation, therefore, is a natural future research direction. Stochastic optimization at a large scale, however, comes with its own challenges, including large computational requirements and long solution times. Therefore, decomposition of the centralized optimization formulation is another compelling near-term research direction, which can help reduce the compute time by parallelizing the architecture. A third future enhancement to the proposed framework is the addition of power flow constraints, such that a cluster of microgrids can be studied to conduct community-level analysis.

With the source code made publicly available, our proposed computational framework for behind-the-meter distributed energy resource techno-economic modeling and optimization offers: (i) transparency—making the mathematical formulation and end-to-end algorithmic framework free to explore; (ii) flexibility—empowering practitioners to make changes to the code to customize the project analysis and answer other interesting research questions; and (iii) collaboration—facilitating collaboration among research, academia, and industry energy modelers to assess and enhance the proposed framework. This framework is made available in three different formats: (i) open-sourced code via a GitHub repository[16]; (ii) Application Programming Interface[17] (API); and (iii) Webtool[18] to enable users of varying levels of expertise and broad professional backgrounds to conduct analysis—with the goal of achieving transformational levels of behind-the-meter DER deployment.

## VII. ACKNOWLEDGMENTS

This work was authored by the National Renewable Energy Laboratory, operated by Alliance for Sustainable Energy, LLC, for the U.S. Department of Energy (DOE) under Contract No. DE-AC36-08GO28308. Funding was provided by the U.S. Department of Energy Office of Energy Efficiency and Renewable Energy Federal Energy Management Program and the Advanced Manufacturing Office. The authors would like to thank Andy Walker, Linda Parkhill, Kathleen Krah, Andrew Jeffery, Nick Muerdter, Xiangkun Li, Rob Eger, and Nicholas DiOrio for help with building and maintaining the REopt Lite API; and Adam Warren for reviewing the manuscript. The authors would also like to thank Rachel Shepherd (Office of Energy Efficiency and Renewable Energy, DOE) and Bob Gemmer (Advanced Manufacturing Office, DOE) for funding this work. The views expressed in the article do not necessarily represent the views of the DOE or the U.S. Government. The U.S. Government retains and the publisher, by accepting the article for publication, acknowledges that the U.S. Government retains a nonexclusive, paid-up, irrevocable, worldwide license to publish or reproduce the published form of this work or allow others to do so, for U.S. Government purposes.

---

[16] https://github.com/NREL/REopt_Lite_API
[17] https://developer.nrel.gov/docs/energy-optimization/reopt-v1/
[18] https://reopt.nrel.gov/tool